\documentclass{article}

\usepackage[english]{babel}

\usepackage[a4paper,top=2cm,bottom=2cm,left=3cm,right=3cm]{geometry}

\usepackage[utf8]{inputenc} 
\usepackage[T1]{fontenc}    
\usepackage[colorlinks=true,citecolor=blue,anchorcolor=purple]{hyperref}
\usepackage{url}            
\usepackage{booktabs}       
\usepackage{amsfonts}       
\usepackage{nicefrac}       
\usepackage{microtype}      
\usepackage[table]{xcolor}         
\usepackage{natbib}
\setcitestyle{authoryear,citesep={;},aysep={,},yysep={;}}
\usepackage{amsmath}
\usepackage{amssymb}
\usepackage{array}
\usepackage{multirow}
\usepackage[font=small]{caption}
\usepackage{subcaption}
\usepackage{graphicx}
\usepackage[ruled,vlined]{algorithm2e}
\usepackage{float}
\usepackage{wrapfig}
\usepackage{adjustbox}
\usepackage{enumitem}

\renewcommand{\cite}{\citep}
\usepackage[english]{babel}
\addto\captionsenglish{%
}
\title{\bf {\tt UniPhyNet}: A Unified Network For Multimodal Physiological Raw Signal Classification}

\author{
  Renxiang Qiu$^{\bot}$, Raghavendra Selvan$^{\bot}$\\[0.5em]
      ${\bot}$~\small{Department of Computer Science, University of Copenhagen, Denmark} \\
  \small{\texttt{kgq725@alumni.ku.dk,raghav@di.ku.dk@di.ku.dk}}
}

\date{}

\begin{document}

\maketitle
\begin{abstract}
We present {\tt UniPhyNet}, a novel neural network architecture to classify cognitive load using multimodal physiological data -- specifically EEG, ECG and EDA signals -- without the explicit need for extracting hand-crafted features. {\tt UniPhyNet} integrates multiscale parallel convolutional blocks and ResNet-type blocks enhanced with channel block attention module to focus on the informative features while a bidirectional gated recurrent unit is used to capture temporal dependencies. This architecture processes and combines signals in both unimodal and multimodal configurations via intermediate fusion of learned feature maps. On the CL-Drive dataset, {\tt UniPhyNet} improves raw signal classification accuracy from 70\% to 80\% (binary) and 62\% to 74\% (ternary), outperforming feature-based models, demonstrating its effectiveness as an end-to-end solution for real-world cognitive state monitoring.

\footnote{Accepted to be presented at the 35th IEEE International Workshop on Machine Learning for Signal Processing (IEEE MLSP 2025)}
\footnote{Source code and appendix are available at : \url{https://github.com/HughYau/UniPhyNet}}
\end{abstract}

\section{Introduction}
Cognitive load measurement is crucial in cognitive science, with significant implications for understanding mental states during complex tasks~\cite{brunken2003direct,mayer2002aids,sweller2011cognitive}. Accurate real-time monitoring of cognitive load is essential in applications such as enhancing driving safety, improving human-computer interaction, and optimizing educational outcomes~\cite{gevins2003neurophysiological,liang2018experimental}. Measurement of cognitive load through physiological signals such as EEG, ECG, and EDA\footnote{EEG: Electroencephalogram, ECG: Electrocardiogram, EDA: Electrodermal activity} can enable intelligent systems to adapt to users' mental states, thereby reducing errors and enhancing performance~\cite{vanneste2021towards,wang2023characterisation,mathur_dynamic_2021}.

Accurately measuring cognitive load is vital in various applications, from educational settings to driver assistance systems. Traditional methods have relied on behavioral metrics and self-reporting, but recent advancements have enabled more objective measurements using physiological data. Pinto et al. (2020) highlighted the improved performance of multimodal approaches combining EEG, ECG, and EDA signals for emotion evaluation~\cite{pinto2020multimodal}, while Wang et al.(2024) proposed multi-source domain generalization techniques to enhance the applicability of physiological signals, such as ECG, for cognitive load estimation in diverse settings, further illustrating the potential of these measures for real-world applications~\cite{wang_multi-source_2024}. Lin and Li (2023) have emphasized the growing trend towards multimodal data fusion for reliable cognitive load measurements~\cite{lin2023review}.


Despite its importance, real-time cognitive load measurement using multimodal physiological data remains complex and under-explored. Traditional research often focuses on unimodal data, which limits the ability to capture the full spectrum of physiological responses associated with cognitive load. Additionally, the scarcity of datasets with comprehensive multimodal physiological data in realistic scenarios has hindered the development of robust models for cognitive load classification.

\begin{figure*}[t]
    \centering
    \includegraphics[width = 0.88\linewidth]{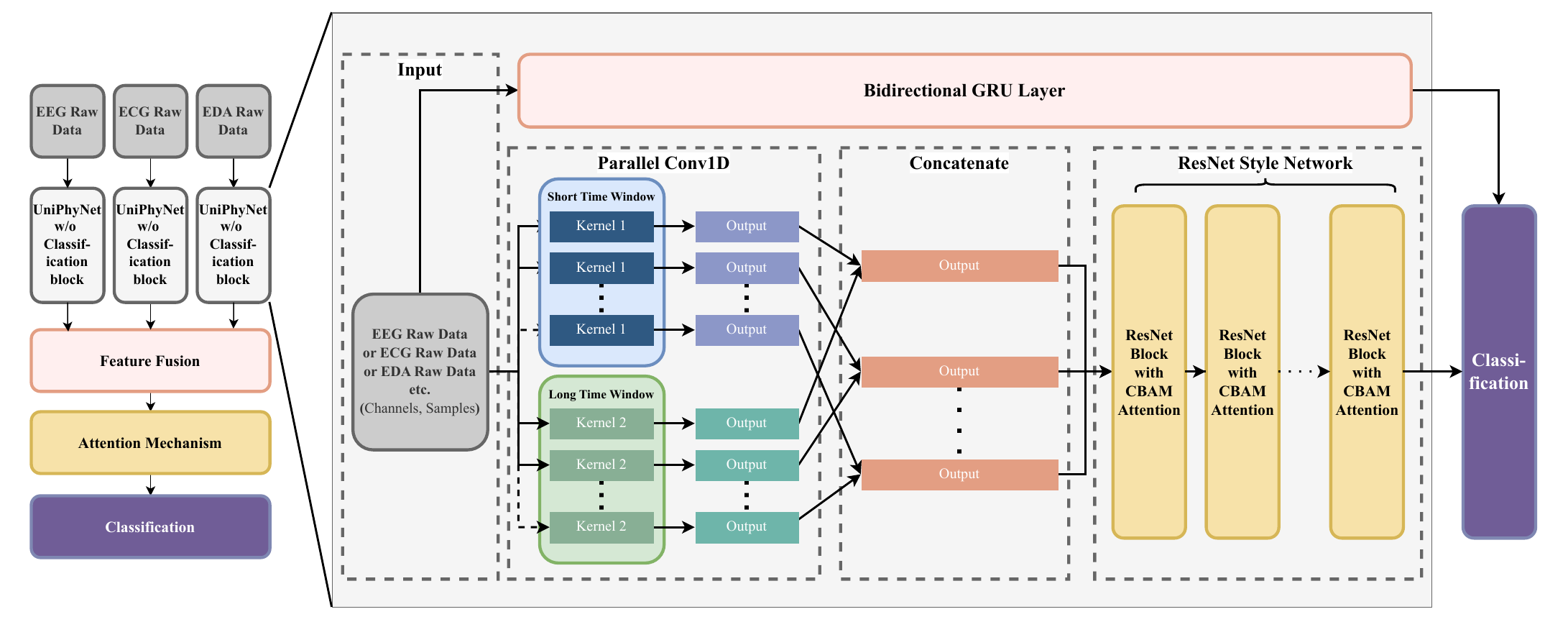}    
    \caption{Overview of the {\tt UniPhyNet} architecture. The multimodal architecture consists of the unimodal blocks which are shown with more fine-grained architectural details.}
    \label{fig:uniphynet}
    \vspace{-4mm}

\end{figure*}
In response to these challenges, recent advances in multimodal physiological data analysis have demonstrated the benefits of integrating multiple signal types to enhance classification accuracy. Multimodal approaches leverage the complementary information provided by different physiological signals, such as EEG, ECG, EDA~\cite{chiossi2024optimizing,ahmad2020framework}, even for EMG~\cite{yan2022emotion,gao_bimodal_2023}, to improve performance over unimodal methods. 


Deep learning methods have further advanced cognitive load and emotion classification. Among the models developed for such tasks, EEGNet stands out as a lightweight convolutional neural network (CNN) designed specifically for EEG signal classification~\cite{lawhern2018eegnet}. Although EEGNet has significantly advanced EEG-based cognitive load classification, its reliance on depth-wise separable convolutions and limited scope for multi-scale feature extraction suggest room for improvement. As for multimodal application, Hssayeni and Ghoraani (2021) proposed a deep learning-based multimodal data fusion framework for affect estimation using ECG, EMG, EDA, and RESP signals \cite{hssayeni2021multi}. Han et al. (2020) developed a multimodal deep learning network to classify pilots' mental states using EEG, ECG, respiration, and EDA data, showcasing the potential of integrated approaches in real-world applications \cite{han2020classification}.


In this work, we introduce {\tt UniPhyNet}, a novel neural network architecture designed to unify the processing of multimodal physiological data. Unlike many models, it operates directly on the preprocessed raw data, eliminating the need for extensive feature extraction. This approach not only simplifies the pipeline, but also improves the ability of the model to generalize across different signal types. {\tt UniPhyNet}'s uniform network structure supports intermediate feature fusion, offering a significant improvement over traditional decision-fusion methods. By eliminating the need for extensive feature engineering, it offers a paradigm shift in cognitive load classification, providing a scalable solution for real-world applications such as adaptive educational technologies and driver monitoring systems.

\section{Methodology}



{\tt UniPhyNet} is implemented in two configurations: a unimodal version for EEG, ECG, or EDA separately, and a multimodal version that integrates features from all data types through intermediate fusion. This dual approach enables comprehensive evaluation, leveraging complementary information across modalities to enhance classification accuracy. Figure~\ref{fig:uniphynet} illustrates the architecture of both unimodal and multimodal version of {\tt UniPhyNet}. 

In the multimodal configuration, the features extracted by the unimodal networks are concatenated and further processed with an attention mechanism to prioritize important features, enhancing the model’s overall performance~\cite{vaswani2017attention}. 

\textbf{1) Parallel Convolutional Blocks:} A significant enhancement in {\tt UniPhyNet} is the incorporation of parallel convolutional blocks, a design inspired by the hierarchical processing of sensory information in the human brain. These blocks leverage convolutional layers with varying kernel sizes (e.g., 3, 5, 9, 11) to capture multiscale temporal features, mimicking the brain's ability to integrate information across multiple scales of time and space. This approach aligns with the theoretical understanding of neural mechanisms, such as temporal receptive fields, that allow the brain to process both short-term and long-term patterns in sensory inputs~\cite{hubel1968receptive}. 

{\tt UniPhyNet} also utilizes the SiLU (Sigmoid Linear Unit) activation function, which provides smoother gradients to improve the learning dynamics and potentially performance~\cite{elfwing2018sigmoid}. The output from the parallel convolutional layers are concatenated along the channel dimension, creating a rich {\em learned} feature set that captures diverse aspects of the input signals.

\textbf{2) ResNet Blocks with Channel Block Attention Module:} {\tt UniPhyNet} enhances feature extraction by replacing EEGNet's traditional depth-wise separable convolutions with ResNet blocks integrated with the Convolutional Block Attention Module (CBAM)~\cite{woo2018cbam}. ResNet blocks utilize residual connections to address the vanishing gradient problem, enabling the training of deeper networks~\cite{he2016deep}. Each block includes batch normalization and ReLU activation before convolution to stabilize training.

The integrated CBAM enhances feature maps by using two attention mechanisms: channel and spatial. Channel attention uses max-pooling and average-pooling to create a channel attention then refines these features to generate a map emphasizing key temporal regions. This dual-attention process teaches the network ‘what’ and ‘where’ to focus on, prioritizing key features across channel and spatial dimensions for optimal data processing~\cite{woo2018cbam}. This approach is based on attention theory, which emphasizes the importance of selective focus for processing high-dimensional inputs~\cite{he_channel_2022}.



\textbf{3) Temporal Dependencies Capture:} {\tt UniPhyNet} also incorporates a bidirectional gated recurrent unit (GRU) layer to capture temporal dependencies in raw input data~\cite{kuanar_cognitive_2018,cho2014learning}. This layer is crucial for modeling temporal relationships, particularly in long-term physiological signal data where time dependencies are significant~\cite{borsdorf_multi-head_2023}. Although existing algorithms like EEGNet are designed to process short-term data segments (typically around 1 second), {\tt UniPhyNet} is capable of handling longer data segments (10 seconds in this study), which allows it to capture the temporal dynamics that evolve over longer periods.

The final fully connected layer in {\tt UniPhyNet} combines features from the combined convolutional output and the GRU layer, resulting in enhanced classification performance. 


\section{Data \& Experiments}
\subsection{CL-Drive Dataset}

We utilized the CL-Drive dataset~\cite{ahmad2020framework}, a comprehensive multimodal dataset specifically designed for cognitive load assessment in driving scenarios. This dataset includes physiological signals such as EEG, ECG, and EDA, collected from 21 participants in an immersive vehicle simulator under various driving conditions designed to induce different levels of cognitive load. The dataset provides rich information, including benchmarks with classical methods, facilitating comparison with our approach\footnote{Dataset is available at \url{https://github.com/prithila05/cl-drive}}.

The dataset consists of four channel EEG data sampled at $256$ Hz, three channel ECG data sampled at $512$ Hz, and three channel EDA data sampled at $128$ Hz. All the three physiological signals were obtained from participants in nine driving scenarios that lasted 3 minutes each and the cognitive load was reported every 10 seconds from very low (1) to very high (9) in the range: $[1,2,...,9]$. 

{\bf Data Preparation}: First, all the data were segmented into 10-second windows corresponding to the cognitive load ratings. Then, to ensure the quality and consistency of the data used for training {\tt UniPhyNet}, several preprocessing steps: For EEG signals, a low-pass filter with a cut-off frequency of $20$ Hz was used to remove low-frequency drifts and high-frequency noise. The signals were then normalized to have zero mean and unit variance, ensuring uniformity across different subjects and sessions. ECG signals underwent band-pass filtering between $0.5$ Hz and $40$ Hz to eliminate baseline wander and high-frequency noise, followed by normalization similar to EEG data. EDA signals were filtered using a band-pass filter with a range of $0.05-3$ Hz to remove high-frequency noise, and subsequently normalized to account for inter-subject variability.


To enhance training set diversity and mitigate overfitting, data augmentation was applied only to the training data within each cross-validation fold with one of three methods: Gaussian noise addition, time warping and amplitude scaling. Additional details on preprocessing and data augmentation are provided in the Appendix.



\subsection{Experimental Set-up}
We train all models using 10-fold, and leave one subject out (LOSO), cross-validation settings. We explore both binary and ternary classification of cognitive load. Certain individuals may not be able to distinguish cognitive load scores to high level of detail. As a result, the scores were converted to ‘binary’ (high/low) and ‘ternary’ (high/medium/low) levels using grouping of the scores. However, this initial coarse labeling scheme allows future research to focus on more detailed classification schemes if necessary. For binary classification, we group the cognitive load ratings from 1 to 4 as ‘low’ cognitive load and 5 to 9 as ‘high’ cognitive load. For ternary classification, we divide cognitive load ratings into 3 groups, 1 to 3, 4 to 6, and 7 to 9, which correspond to classes of cognitive load 'low', 'medium' and 'high', respectively.

{\bf Benchmarking}: The CL-Drive dataset includes benchmark classification results for various machine learning and deep learning models on both binary and ternary cognitive load classification tasks. For classical machine learning: they train a total of 9 machine learning classifiers namely AdaBoost (AB), Decision Tree (DT), Naive Bayes (NB), K-Nearest Neighbor (KNN), Linear Discriminant Analysis (LDA), Random Forest (RF), Support Vector Machine (SVM), Extreme Gradient Boosting (XGB), and Multi-Layer Perceptron (MLP).  For deep learning network they use two deep CNNS, a VGG-style network, and a ResNet-style network. The VGG-style network has two main blocks, where each block consists of two Conv1D layers, batch normalization, ReLU activation, and maximum pooling operation. These blocks are followed by two fully connected layers and a classification layer. Cross-entropy loss with a learning rate of $0.001$ was used for training. ADAM was used as an optimizer for their network.


\begin{figure}[t]
    \centering
    \includegraphics[width=\linewidth]{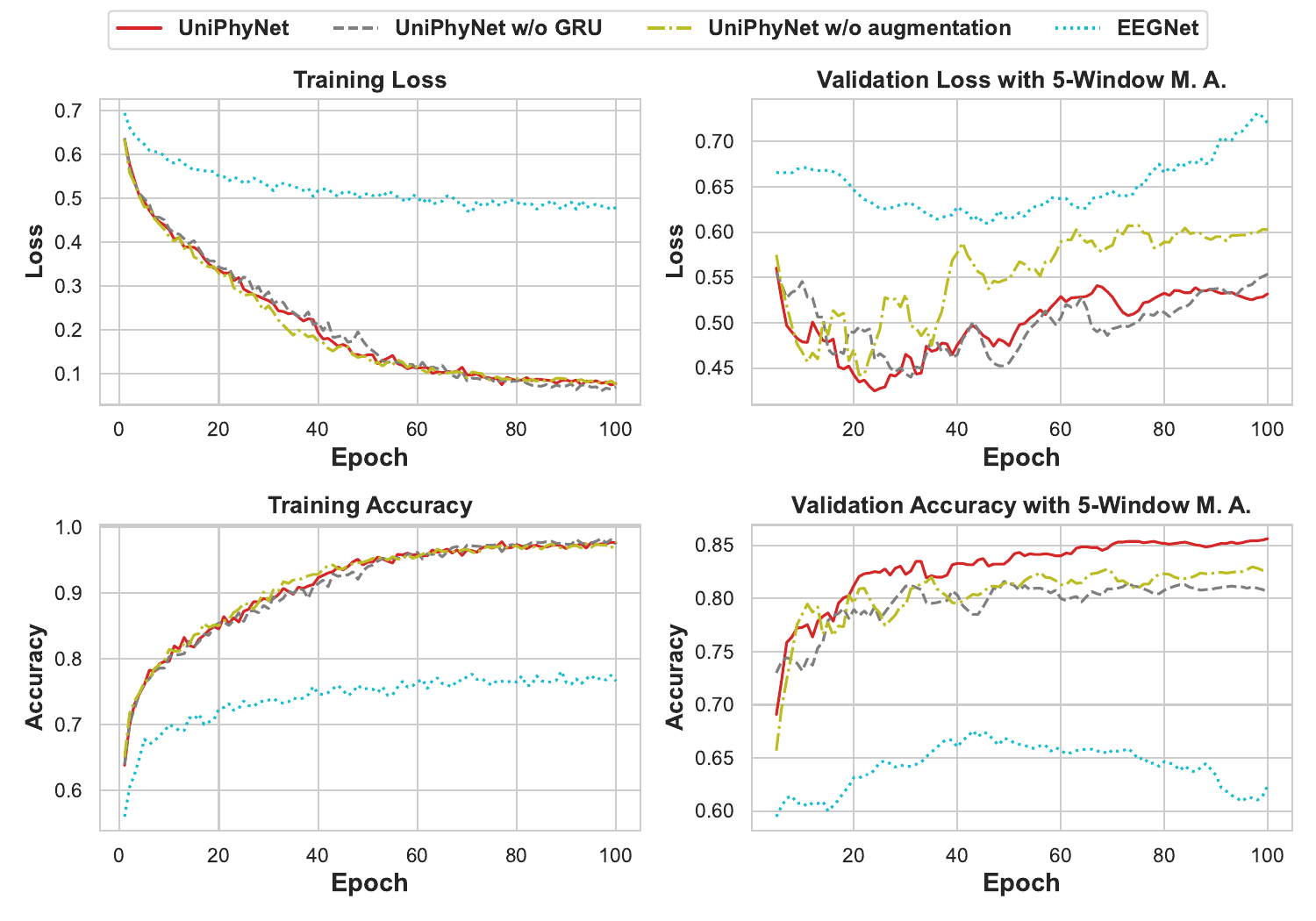}
    \caption{Comparison of {\tt UniPhyNet} in ablation settings.}
    \label{fig:comp1}
    \vspace{-4mm}
\end{figure}
\begin{figure}[t]
    \centering
    \includegraphics[width=\linewidth]{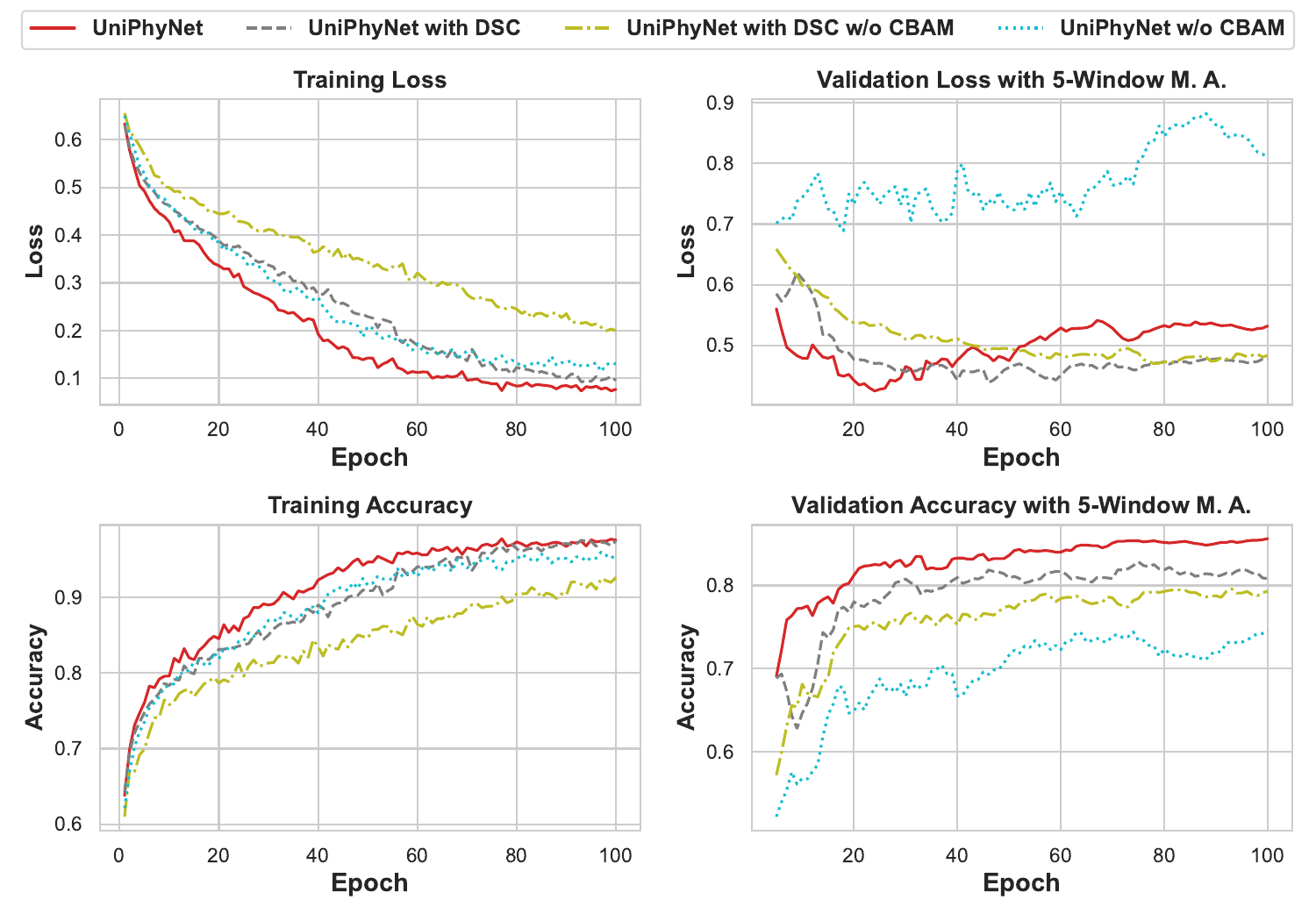}
    \caption{Effect of CBAM module in {\tt UniPhyNet}.}
    \label{fig:comp2}
    \vspace{-4mm}
\end{figure}
\begin{figure}[t]
    \centering
    \includegraphics[width=\linewidth]{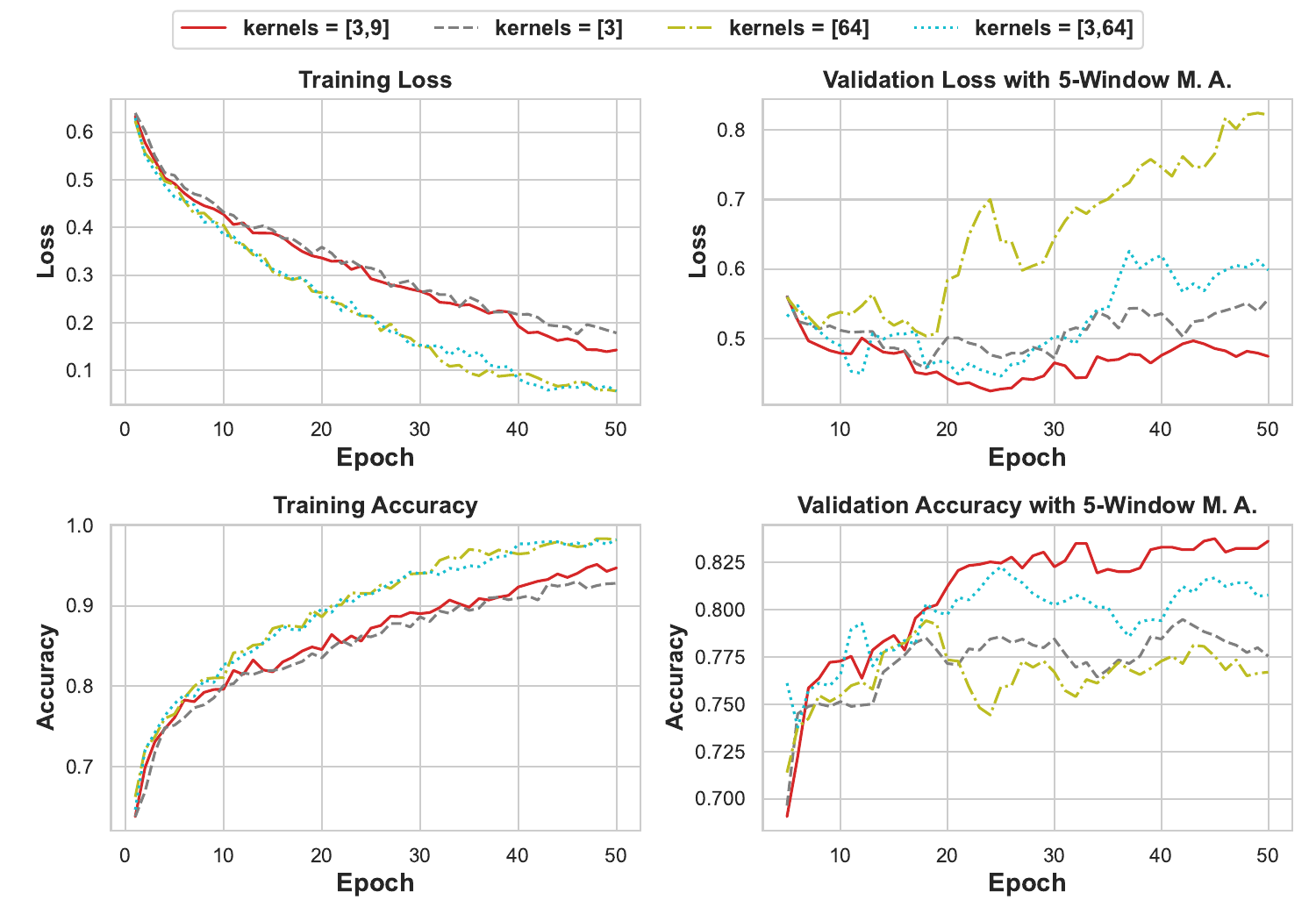}
    \caption{Effect of different kernels in {\tt UniPhyNet}.}
    \label{fig:comp3}
    \vspace{-4mm}
\end{figure}

\subsection{Ablation Study}

To assess the effectiveness of our {\tt UniPhyNet} model, we performed extensive evaluations using the CL-Drive dataset, specifically for binary classification tasks on unimodal EEG data. The dataset was randomly divided, with 90\% used for training and the remaining 10\% for validation. The following parameters were applied: {\tt UniPhyNet} models used kernels set to $[3,9]$, with 64 feature maps and 8 ResNet blocks. For comparison, the EEGNet model was configured with a kernel length of 128 (as suggested by the original authors) and a learning rate of $0.001$. A learning rate scheduler, ReduceLROnPlateau, was employed to adaptively reduce the learning rate based on validation loss, with a reduction factor of 0.5 and patience of 15 epochs. The experiments were performed on a standard laptop with an Nvidia RTX4060 GPU and Intel i5-12450H processor.

Initially, we compared {\tt UniPhyNet} with EEGNet and two ablation versions of {\tt UniPhyNet} -- one without the GRU layer and one without data augmentation. The results, shown in Figure \ref{fig:comp1}, indicate that {\tt UniPhyNet} outperforms EEGNet on both training and validation datasets. While the differences in training performance between {\tt UniPhyNet} and its ablation models are minimal, {\tt UniPhyNet} shows significantly better performance on the validation set. This highlights the importance of the GRU layer and data augmentation in enhancing model generalization and reducing overfitting.

Next, we explored the components of {\tt UniPhyNet}'s ResNet blocks to evaluate the effectiveness of ResNet and CBAM attention mechanisms. Specifically, we compared models using Depth Separable Convolution (DSC) layers, which are core to EEGNet, with and without the CBAM mechanism in the ResNet blocks. As shown in Figure \ref{fig:comp2}, incorporating ResNet blocks with CBAM attention mechanisms significantly enhances performance, especially in validation metrics. Although both DSC and CBAM improve the model’s performance, CBAM has a more substantial impact. Notably, combining DSC with CBAM does not further enhance performance beyond what CBAM alone achieves, indicating that CBAM in ResNet blocks is the most effective strategy for improving {\tt UniPhyNet}'s performance on multimodal physiological signal classification.

The final evaluation examined the impact of different kernel sizes in {\tt UniPhyNet}'s parallel convolution layers. Four configurations were tested: $[3,9], [3], [64],$ and $[3,64]$. As depicted in Figure \ref{fig:comp3}, larger kernels like $[64]$ resulted in quicker convergence during training but also led to a significant increase in validation loss, indicating a higher risk of overfitting. In contrast, combining smaller and larger-sized kernels, as in the $[3,64]$ configuration, mitigated this risk. The $[3,9]$ configuration maintained lower validation loss and higher validation accuracy over time, suggesting that a combination of smaller and medium-sized kernels effectively captures a diverse set of features, reducing overfitting while maintaining generalization. The $[3]$ configuration performed reasonably well but did not achieve the same level of generalization as $[3,9]$.

\begin{table}[h]
\centering
\scriptsize
\caption{Accuracy (\& F1 score) in 10-fold binary setup.}
\resizebox{\columnwidth}{!}{%
\begin{tabular}{@{}ccccc@{}}
\toprule
\textbf{Models} & \textbf{EEG}           & \textbf{EEG, ECG}      & \textbf{EEG, EDA}      & \textbf{EEG, ECG, EDA} \\ \midrule
AB              & 67.17 (55.37)          & 73.26 (68.43)          & 71.09 (66.53)          & 73.84 (69.85)          \\
DT              & 65.31 (63.04)          & 72.77 (71.15)          & 69.61 (67.93)          & 73.02 (71.41)          \\
NB              & 48.68 (46.54)          & 51.80 (50.48)          & 51.39 (49.97)          & 53.73 (52.87)          \\
KNN             & 70.61 (68.49)          & 69.34 (67.40)          & 71.78 (70.16)          & 70.64 (69.16)          \\
LDA             & 66.83 (62.45)          & 72.74 (70.65)          & 71.19 (68.45)          & 74.73 (72.94)          \\
RF              & \underline{ 77.41 (73.39)}    & \underline{ 79.34} (76.27)          & \underline{ 79.48 (76.47)}    & \underline{ 81.26} (78.81)    \\
SVM             & 61.88 (38.29)          & 62.08 (38.89)          & 61.88 (38.29)          & 64.35 (46.59)          \\
XGB             & 77.38 (73.72)          & \textbf{82.95 (81.25)} & \textbf{80.06 (77.67)} & \textbf{82.61 (80.94)} \\ \midrule
MLP             & 74.32 (72.36)          & 74.22 (72.31)          & 76.31 (74.02)          & 76.00 (74.54)          \\
VGG (feat.)     & 75.56 (73.21)          & 77.57 (75.80)          & 78.99 (76.94)          & 78.78 (77.22)          \\
ResNet (feat.)  & 69.38 (65.26)          & 74.27 (71.48)          & 71.74 (68.46)          & 75.49 (72.71)          \\
VGG (raw)       & 63.83 (63.23)          & 67.73 (66.97)          & 66.95 (66.11)          & 70.12 (69.20)          \\
ResNet (raw)    & 61.95 (59.75)          & 64.49 (62.14)          & 60.90 (57.45)          & 64.41 (62.82)          \\ \midrule
{\tt UniPhyNet}     & \textbf{79.29 (79.31)} &  79.33 \underline{ (79.24)}    & 77.38(77.34)           & 80.16 \underline{(80.25)}          \\ \bottomrule
\end{tabular}%
}
\label{tab:res1}
\end{table}

\begin{table}[h]
\centering
\scriptsize
\caption{Accuracy (\& F1 score) in LOSO binary setup.}
\label{tab:res2}
\resizebox{\columnwidth}{!}{%
\begin{tabular}{@{}ccccc@{}}
\toprule
\textbf{Models} & \textbf{EEG}          & \textbf{EEG, ECG}      & \textbf{EEG, EDA}      & \textbf{EEG, ECG, EDA} \\ \midrule
AB              & 62.30 (46.81)         & 66.58 (59.47)          & 63.01 (54.57)          & 67.86 (62.22)          \\
DT              & 54.63 (49.73)         & 60.33 (54.82)          & 57.94 (53.07)          & 60.97 (56.91)          \\
NB              & 47.80 (43.54)         & 48.94 (45.80)          & 48.16 (44.71)          & 49.85 (47.52)          \\
KNN             & 58.21 (53.11)         & 61.45 (58.09)          & 60.83 (56.10)          & 62.51 (59.51)          \\
LDA             & 57.06 (49.61)         & 59.87 (55.67)          & 62.95 (56.99)          & 63.15 (58.60)          \\
RF              & 63.82 (50.97)         & 65.76 (56.84)          & 66.64 (56.73)          & 67.95 (59.92)          \\
SVM             & 62.01 (37.65)         & 59.85 (37.84)          & 61.80 (37.91)          & 61.48 (45.71)          \\
XGB             & 62.98 (52.34)         & 66.61 (60.53)          & 66.39 (59.34)          & 69.37 (64.01)          \\ \midrule
MLP             & 57.86 (51.98)         & 63.64 (57.90)          & 63.44 (58.13)          & 64.48 (60.33)          \\
VGG (feat.)     & \underline{ 70.70 (64.22)}   & \underline{ 74.72} (70.68)          & \textbf{73.01} \underline{ (68.08)} & \textbf{76.17} \underline{(71.72)} \\
ResNet (feat.)  & 67.45 (61.39)         & \textbf{75.90} \underline{ (71.62)} & \underline{ 72.20} (66.48)          & \underline{ 74.23} (69.28)          \\
VGG (raw)       & 65.00 (58.92)         & 63.67 (57.59)          & 67.18 (60.78)          & 67.37 (62.71)          \\
ResNet (raw)    & 65.79 (57.58)         & 63.99 (56.73)          & 69.03 (61.67)          & 68.74 (61.88)          \\ \midrule
{\tt UniPhyNet}       & \textbf{73.47(73.61)} & 73.61\textbf{(74.06)}     &  70.09 \textbf{(70.38)}  & 73.94 \textbf{(74.59) }    \\ \bottomrule
\end{tabular}%
}
\end{table}

\section{Results}

We present the results on the CL-Drive dataset using 10-fold and LOSO cross validation in the binary classification task to compare {\tt UniPhyNet} with various benchmark models from the dataset. In these tables, bold values denote the highest, while underline represents the second-highest. The models were evaluated for classification tasks across different data modalities (EEG, ECG, and EDA). The model configurations are: For EEG Data: kernels = $[3, 9]$, ResNet block number = 8; For ECG data: kernels = $[5, 11]$, ResNet block number = 9 (due to the higher frequency of $512$ Hz, need deeper network to reduce to the same feature numbers as EEG data for balance); for EDA data: kernels = $[13]$, ResNet block number = 7 (due to the lower frequency of $128$ Hz). Models for all the data have the same number of feature maps: 64 for unimodal, 32 for multimodal due to GPU memory constraints.

The results reported in Table~\ref{tab:res1} and Table~\ref{tab:res2} show that {\tt UniPhyNet} outperforms all benchmark models in the binary tasks for unimodal EEG. This demonstrates the effectiveness of {\tt UniPhyNet}'s architecture in handling EEG data. For multimodal data (EEG, ECG, EDA), {\tt UniPhyNet} achieves performance comparable to the best performing models in binary classification tasks, which often use decision trees with pre-extracted features.

\begin{table}[h]
\centering
\caption{Accuracy (\& F1 score) in 10-fold ternary setup.}
\scriptsize
\label{tab:res3}
\resizebox{\columnwidth}{!}{%
\begin{tabular}{@{}ccccc@{}}
\toprule
\textbf{Models} & \textbf{EEG}           & \textbf{EEG, ECG}      & \textbf{EEG, EDA}      & \textbf{EEG, ECG, EDA} \\ \midrule
AB              & 46.20 (38.35)          & 51.84 (48.65)          & 51.56 (48.78)          & 53.52 (51.62)          \\
DT              & 48.95 (48.77)          & 56.03 (56.08)          & 52.08 (51.85)          & 57.48 (57.36)          \\
NB              & 34.93 (29.77)          & 37.13 (33.00)          & 37.47 (33.77)          & 39.50 (36.55)          \\
KNN             & 54.59 (53.95)          & 51.60 (50.81)          & 56.34 (55.90)          & 52.46 (51.67)          \\
LDA             & 53.01 (51.93)          & 58.27 (58.09)          & 58.13 (57.82)          & 61.02 (61.00)          \\
RF              & 63.56 (63.04)          & 68.41 (68.42)          & \underline{ 69.34 (69.17)}    & 71.57 (71.63)          \\
SVM             & 41.01 (21.66)          & 46.58 (37.78)          & 41.73 (24.17)          & 48.40 (41.25)          \\
XGB             & \underline{ 64.49 (64.14)}    & \underline{ 70.78 (71.01)}    & \textbf{71.74 (71.69)} & \underline{ 73.50 (73.76)}    \\ \midrule
MLP             & 58.44 (57.72)          & 61.50 (60.64)          & 61.12 (60.90)          & 62.80 (62.66)          \\
VGG (feat.)     & 62.12 (60.92)          & 62.85 (62.21)          & 64.44 (63.91)          & 65.76 (65.37)          \\
ResNet (feat.)  & 47.19 (44.61)          & 55.52 (53.74)          & 51.91 (50.66)          & 56.15 (55.48)          \\
VGG (raw)       & 47.85 (43.7)           & 55.43 (51.37)          & 56.48 (51.61)          & 61.76 (57.96)          \\
ResNet (raw)    & 50.82 (37.41)          & 56.56 (50.09)          & 53.67 (44.25)          & 60.62 (54.07)          \\ \midrule
{\tt UniPhyNet}       & \textbf{68.24 (68.13)} & \textbf{73.60 (73.78)} & 67.36 (67.35)          & \textbf{74.13 (74.12)} \\ \bottomrule
\end{tabular}%
}
\end{table}

\begin{table}[h]
\centering
\scriptsize
\caption{Accuracy (\& F1 score) in LOSO ternary setup.}
\resizebox{\columnwidth}{!}{%
\begin{tabular}{@{}ccccc@{}}
\toprule
\textbf{Models} & \textbf{EEG}          & \textbf{EEG, ECG}     & \textbf{EEG, EDA}      & \textbf{EEG, ECG, EDA} \\ \midrule
AB              & 37.79 (27.09)         & 42.13 (34.39)         & 42.56 (36.04)          & 44.26 (38.46)          \\
DT              & 35.83 (33.35)         & 40.15 (37.63)         & 37.37 (34.68)          & 35.77 (33.02)          \\
NB              & 33.28 (26.3)          & 33.81 (27.37)         & 33.02 (27.85)          & 33.23 (28.40)          \\
KNN             & 35.19 (32.87)         & 39.19 (37.06)         & 37.40 (34.97)          & 39.24 (37.31)          \\
LDA             & 36.18 (33.64)         & 40.23 (37.02)         & 40.33 (37.61)          & 42.31 (39.07)          \\
RF              & 37.02 (32.58)         & 40.15 (37.54)         & 39.97 (36.05)          & 42.28 (40.15)          \\
SVM             & 38.48 (20.41)         & 39.66 (31.73)         & 38.42 (21.00)          & 39.83 (33.22)          \\
XGB             & 36.09 (32.83)         & 40.06 (38.11)         & 41.63 (38.50)          & 44.48 (42.31)          \\ \midrule
MLP             & 38.30 (34.11)         & 39.30 (36.70)         & 41.44 (36.83)          & 42.04 (38.88)          \\
VGG (feat.)     & 49.21 (43.75)         & 54.34 (48.93)         & 51.83 \underline{(47.64)}          & 54.39 \underline{(50.96)}          \\
ResNet (feat.)  & 47.30 (42.08)         & 52.82 \underline{(49.44)}         & 53.44 (47.1)           & 55.31 (50.65)          \\
VGG (raw)       & 57.91 (44.12)         & \underline{ 60.84} (47.61)   & \textbf{58.86} (44.83) & \underline{ 61.29} (49.18)    \\
ResNet (raw)    & \underline{ 58.13 (45.24)}   & 60.36 (47.58)         & \underline{ 58.22} (43.81)    & 60.12 (46.68)          \\ \midrule
{\tt UniPhyNet}       & \textbf{60.45(59.43)} & \textbf{62.64(62.02)} & 57.98 \textbf{(57.70)}           & \textbf{63.76(63.58)}  \\ \bottomrule
\end{tabular}%
}
\label{tab:res4}
\end{table}

We also report the results from experiments for ternary classification in Table~\ref{tab:res3} and Table~\ref{tab:res4}. In ternary tasks, {\tt UniPhyNet} demonstrates superior performance compared to traditional feature-based models, achieving a significant increase in classification accuracy, highlighting its robustness in handling complex tasks and multimodal data.

This improvement highlights the advantages of end-to-end learning frameworks in leveraging the full informational content of raw data, bypassing the potential biases and limitations of manually engineered features. From a theoretical perspective, this aligns with the premise of representation learning, where models are trained to discover underlying patterns directly from data, a concept foundational to deep learning's success~\cite{lecun2015deep}. 

When compared with deep learning models like VGG and ResNet that process raw data, {\tt UniPhyNet} consistently delivers better performance. This emphasizes the advantage of our tailored deep network structure for physiological signal classification, particularly in ternary tasks and multimodal scenarios. 
\begin{table}[h]
\centering
\scriptsize
\caption{Accuracy (\& F1 score) for {\tt UniPhyNet} with ECG/ EDA data.}
\begin{tabular}{@{}ccc@{}}
\toprule
                 & \textbf{ECG}  & \textbf{EDA}  \\ \midrule
\textbf{Binary}  & 78.64 (78.61) & 66.30 (66.24) \\
\textbf{Ternary} & 71.25 (71.30) & 50.29 (42.76) \\ \bottomrule
\end{tabular}
\label{tab:ecg_eda}
\end{table}

In addition, we also evaluated the classification task performance using only ECG and EDA data using 10-fold cross validation, which did not have a benchmark on CL-Drive reported in Table~\ref{tab:ecg_eda}. {\tt UniPhyNet} maintains high classification performance with ECG data, demonstrating its ability to generalize across different physiological signals. However, the performance on EDA data is lower, possibly due to hyperparameter settings and the inherent characteristics of EDA signals related to cognitive load.

\textit{Limitations:}  The proposed {\tt UniPhyNet} is more complex than the other deep learning models compared in this work. It requires more compute than the simpler baseline methods. However, as {\tt UniPhyNet} can process raw physiological data (without feature extraction), and the large performance gains, these additional costs can be justified.

\section{Conclusion}

In this work we presented -- {\tt UniPhyNet} -- a novel architecture for classification of raw physiological signals from multimodal sources. The architecture is based on combining modules that operate on unimodal data and fusing the unimodal representations using channel block attention mechanism. We have shown with comprehensive evaluation on the CL-Drive dataset that {\tt UniPhyNet} is able to utilize the multimodal information to a better extent than the baseline models.

\newpage

\bibliography{references}
\bibliographystyle{abbrvnat}

\newpage

\appendix
\appendix

\label{sec:appendix_a}

\section{Data Preprocessing and Augmentation Pipeline}
\label{subsec:appendix_preprocessing}

\subsection{EEG Filtering and Artifact Handling:}
The EEG signals were preprocessed to isolate cognitively relevant information and reduce noise.
\begin{itemize}
    \item \textbf{Filtering:} We chose a band-pass filter from 0.5 Hz to 40 Hz. This range is standard in cognitive neuroscience as it includes delta (0.5-4 Hz), theta (4-8 Hz), alpha (8-13 Hz), and beta (13-30 Hz) bands, all of which are strongly associated with cognitive processes, including workload. Frequencies above 40 Hz are often contaminated by muscle artifacts (EMG), while those below 0.5 Hz are typically slow drifts.
    \item \textbf{Artifacts:} While techniques like Independent Component Analysis (ICA) can remove eye-blink artifacts, we did not apply them explicitly. Our end-to-end approach, particularly with the CBAM attention mechanism, is designed to be robust to such artifacts. The model can learn to down-weight the temporal segments or channels most affected by noise, effectively ignoring them during classification. Power-line interference was handled using a 60 Hz notch filter as described in the original CL-Drive dataset methodology.
\end{itemize}

\subsection{Data Augmentation Parameters}
To enhance the diversity of the training set and improve model generalization, data augmentation was applied stochastically to the training data within each cross-validation fold. The augmentation was applied uniformly across all classes. The following transformations were used:

\begin{itemize}
    \item \textbf{Gaussian Noise Addition:} White Gaussian noise with a standard deviation of $\sigma = 0.02$ relative to the signal's standard deviation was added to each sample. This simulates minor sensor noise and enhances robustness.
    \item \textbf{Time Warping:} The time axis of the signal was warped using a smooth curve defined by cubic splines. The magnitude of the warp was randomly chosen, with a maximum warp factor of $\pm10\%$. This simulates natural variations in the timing of physiological responses.
    \item \textbf{Amplitude Scaling:} The entire signal was scaled by a random factor drawn from a uniform distribution between 0.8 and 1.2. This accounts for inter-trial variations in signal strength and electrode impedance.
\end{itemize}
These parameters were selected empirically to introduce meaningful variability without distorting the underlying physiological characteristics of the signals.

\section{Model Training and Architectural Rationale}
\label{subsec:appendix_training}

\subsection{Training Hyperparameters}
To ensure full reproducibility, Table \ref{tab:hyperparameters} details the training hyperparameters used for all classical and deep learning models benchmarked in this study. The parameters for the baseline models are sourced from the original CL-Drive publication.

\begin{table*}[h]
\centering
\caption{Training Hyperparameters for All Models.}
\label{tab:hyperparameters}
\resizebox{\textwidth}{!}{%
\begin{tabular}{l|l|l|l|l|l|l}
\hline
\textbf{Model} & \textbf{Learning Rate} & \textbf{Optimizer} & \textbf{Batch Size} & \textbf{Epochs} & \textbf{Regularization} & \textbf{Other Key Parameters} \\
\hline
\multicolumn{7}{c}{\textbf{Classical Machine Learning Models}} \\
\hline
AdaBoost (AB) & 0.1 & - & - & - & - & n\_estimators: 70, algorithm: SAMME.R \\
Decision Tree (DT) & - & - & - & - & - & criterion: gini, max\_depth: 3 \\
Naive Bayes (NB) & - & - & - & - & - & var\_smoothing: $1e^{-9}$ \\
KNN & - & - & - & - & - & n\_neighbors: 20, weights: distance \\
LDA & - & - & - & - & - & solver: lsqr \\
Random Forest (RF) & - & - & - & - & - & n\_estimators: 1000, max\_depth: 50 \\
SVM & - & - & - & - & C: 0.1 & kernel: polynomial \\
XGBoost (XGB) & 0.001 & - & - & - & lambda: 0.0001 & n\_estimators: 1000, max\_depth: 20 \\
MLP & adaptive & - & - & 1000 & - & hidden\_layers: (100, 50) \\
\hline
\multicolumn{7}{c}{\textbf{Deep Learning Baseline Models}} \\
\hline
VGG (feat.) & 0.001 & Adam & 32 & - & Dropout: 0.25 & 2 FC layers \\
ResNet (feat.) & 0.01 & Adam & 32 & - & - & 3 FC layers \\
VGG (raw) & 0.001 & Adam & 256 & - & Dropout: 0.25 & 3 Conv Blocks, 2 FC layers \\
ResNet (raw) & 0.01 & Adam & 256 & - & Dropout: 0.5 & 3 Conv Blocks, 3 FC layers \\
\hline
\multicolumn{7}{c}{\textbf{Proposed Model (UniPhyNet)}} \\
\hline
UniPhyNet & 0.001 & AdamW & 64 & 100 & Weight decay: 1e-4 & ReduceLROnPlateau (factor=0.5, patience=15) \\
\hline
\end{tabular}%
}
\end{table*}

\subsection{Rationale for Architectural Parameter Choices}
The architecture of UniPhyNet was designed with modality-specific parameters to effectively process heterogeneous signals before fusion.

\begin{itemize}
    \item \textbf{Kernel Sizes:} The parallel convolutional kernels were chosen based on signal characteristics and sampling rates. For ECG (512 Hz), larger kernels of [5,11] were used to capture the morphology of the QRS complex, which is sharp and brief. For EEG (256 Hz), a combination of smaller and medium kernels [3,9] was chosen to capture both fast oscillations and slower wave patterns. For EDA (128 Hz), a single larger kernel of [13] was sufficient, as EDA is a much slower-changing signal where broader temporal trends are more informative.
    
    \item \textbf{Number of ResNet Blocks:} The number of ResNet blocks was chosen to progressively downsample the temporal dimension of the feature maps, ensuring that the sequence lengths from all modalities were comparable before the final concatenation and classification stage. Since ECG has the highest sampling rate (512 Hz), it required the most blocks (9) to reduce its sequence length. EEG (256 Hz) required 8 blocks, and EDA (128 Hz) required the fewest (7).
    
    \item \textbf{Feature Map Counts:} The decision to use 64 feature maps for unimodal models and 32 for multimodal models was an empirical choice driven by a trade-off between model capacity and computational constraints. The experiments were conducted on a single GPU with limited memory (Nvidia RTX4060). The multimodal model processes up to three data streams in parallel, which significantly increases memory requirements. Reducing the feature maps to 32 per stream was necessary to allow the model to train without exceeding available GPU memory, while still providing sufficient capacity for effective feature learning.
\end{itemize}

\begin{table}[h!]
\centering
\caption{UniPhyNet Hyperparameters per Modality.}
\label{tab:hyperparams}
\begin{tabular}{lccc}
\toprule
\textbf{Parameter} & \textbf{EEG} & \textbf{ECG} & \textbf{EDA} \\
\midrule
Sampling Rate (Hz) & 256 & 512 & 128 \\
Parallel Kernels & {[}3, 9{]} & {[}5, 11{]} & {[}13{]} \\
ResNet Blocks & 8 & 9 & 7 \\
Feature Maps (Unimodal) & 64 & 64 & 64 \\
Feature Maps (Multimodal) & 32 & 32 & 32 \\
\bottomrule
\end{tabular}
\end{table}

\section{C. Experimental Protocol and Evaluation}
\label{sec:appendix_exp}

\subsection{Leave-One-Subject-Out (LOSO) Cross-Validation:}
The LOSO protocol was implemented to assess the model's ability to generalize to unseen subjects. The procedure is as follows: for a dataset with $N=21$ subjects, the model is trained on data from $N-1$ subjects and tested on the data of the single held-out subject. This process is repeated $N$ times, ensuring each subject serves as the test set exactly once. The final performance metrics reported in Tables 2 and 4 are the \textbf{average accuracy and F1-score across all 21 folds}. We will include standard deviations in the final version of the paper to reflect inter-subject variability. Data augmentation was strictly performed \textit{after} the LOSO split, on the training data of each fold only.

\subsection{Train-Test Split Rationale:}
The 90\%-10\% train-validation split used for the ablation study (Section 3.3) was chosen to maximize the data available for training. Since the purpose of the ablation study was to compare different architectural components, a larger training set allows the model's capabilities to be more fully expressed, making the performance differences between configurations more apparent. While other splits like 80-20 are common, our choice was tailored to the specific goal of that experiment.



\end{document}